\documentclass{article}
\usepackage[final]{graphics}

\newcommand{\be}{\begin{equation}}
\newcommand{\ee}{\end{equation}}
\newcommand{\ba}{\begin{eqnarray}}
\newcommand{\ea}{\end{eqnarray}}

\usepackage{psfig}

\begin{document}
\title{Polarization enhancement in $\vec{d}(\vec{p}$,$^2$He)n  reaction:
Nuclear teleportation}
\maketitle
\begin{center}
S. Hamieh\\[0.3cm]
{\it Kernfysisch Versneller Instituut,\\
Zernikelaan
25,
9747 AA Groningen,
The Netherlands.}
\end{center}
\begin{abstract}
%The purpose of this report is to show that experimental setups and instruments developed for
% conventional nuclear-physics studies allow one to design a new way of performing
%massive  matter teleportation, with a prospect to implement the project in a rather
%{\it short time}.
I show that an experimental technique used in nuclear physics
 may
be successfully applied to quantum teleportation (QT) of spin states of massive  matter. A {\it new} non-local
physical effect the `quantum-teleportation-effect' is discovered for the nuclear polarization measurement.
Enhancement of the neutron polarization is expected in the proposed experiment for QT that
discriminates {\it only} one of the Bell states.
\end{abstract}
\section{Introduction}

Discovery of
 quantum teleportation (QT) \cite{benn} is one of the most profound results
of quantum information theory. By means of a classical communication
 channel and a quantum source realized by a non-local entangled state
such as an EPR-pair of particles, the teleportation process allows to transmit
an unknown quantum state from a sender to a receiver
 which are spatially separated. Experimental realizations of QT have so far been limited
to teleportation of light rather than massive fermion \cite{1,2}. The present work gives an experimental
scheme for QT of proton spin to neutron spin in $\vec{d}(\vec{p}$,$^2$He)n reaction.

%%%\begin{figure}[tb]
%\vspace*{-3.1cm}
%\hspace*{1.cm}
%%%\centerline{\psfig
%%%{width=10cm,figure=teleportation1.eps}}
%%%\caption{ Layout of experiments on proton teleportation.
%%% \protect\label{teleportation}
%%%}
%%%\end{figure}

%When the interaction is terminated and the particles fled
%far away from each other they are as yet described by the same wave function. However
%individual states of each separated particle are completely unknown. Moreover, definite
%individual properties do not exist in principle as the QM postulates

\section{QT with  $\vec{d}(\vec{p}$,$^2$He)n  reaction}
%The proposed experiment for  teleportation is $\vec{d}(\vec{p}$,$^2$He)n reaction.
Preparing a polarized $\vec{d}$ target
 $^3S_1$ and $m_s=0$ state
  with polarized proton bean and selecting knockout reaction,
will lead to the QT of the beam polarization  to the
outgoing neutron. In fact, for clarity reason two assumptions are made:

{\it Assumption I:} assuming the initial states of the deuteron target and the proton beam
are pure\footnote{
%genralization to  mixed ensemble is straightforward. The
General procedure
of QT of mixed state through noisy channel could be found in Ref. \cite{San1}.},
under such condition, the deuteron spin state could be written as
\be |\psi^+\rangle_{23}={1\over \sqrt{2}}(|01\rangle+|10\rangle)\,,\ee
and the proton spin state as
\be  |\Phi\rangle_{1}=a|0\rangle+b|1\rangle\,,\ee
where 1, 2, 3 denote the two protons, and the neutron states, respectively,
and $|0\rangle$, $|1\rangle$ correspond to the positive, negative projection of the spin
to the quantization axis, respectively.
Using the so-called
Bell's basis such that
\be |\psi^{\pm}\rangle={1\over \sqrt{2}}(|01\rangle\pm|10\rangle)\,;
\quad |\phi^{\pm}\rangle={1\over \sqrt{2}}(|00\rangle\pm|11\rangle)\,,\ee
the initial spin state of the system ($\vec{d}$, $\vec{p}$) could be written as follows

 \ba|\Psi\rangle_{123}&=&\frac{1}{2}\left\{|\phi^+\rangle_{12}(a|1\rangle+b|0\rangle)_3
+|\phi^-\rangle_{12}(a|1\rangle-b|0\rangle)_3+|\psi^+\rangle_{12}(a|0\rangle+b|1\rangle)_3\right.\nonumber\\
&&
\left.+|\psi^-\rangle_{12}(a|0\rangle-b|1\rangle)_3 \right\}\label{123}
\,.\ea
%Because the recoil `diproton' is unbound, this reaction leads to a continuum
%of neutron energies. However, because the two proton undergo strong S state
%final state interaction when their relative energy is low, a strong peak occurs at the
%highest neutron energy. Also, at low pp relative energy ( i.e. high neutron energy),
%the pauli principle restricts the two protons to be antiparallel ($^1\rm S_0$ state).
%Hence

{\it Assumption II}: assuming that the polarization transfer and the induced polarization at $0^{\circ}$ are neglected
for knockout reaction. This assumption
is supported by  the low measured $K_y^{y'}\sim -0.1$ \cite{Ohls72} of the
reaction  $1+\vec{{1\over 2}}\rightarrow \vec{{1\over 2}}+0$  that has the  same spin
structure of $\vec{d}(\vec{p}$,$^2$He)n  reaction for unpolarized $d$. This
suggests that  the $\vec{1}+\vec{{1\over 2}}\rightarrow \vec{{1\over 2}}+0$ will have low  $K_y^{y'}$.
Therefore, measuring the polarization of the high energy  neutron, i.e. low pp relative energy
 will force the two outgoing proton state to be singlet and project
 the wave function of  Eq. \ref{123} after measurement into \cite{Ohls72}

 \ba|\Psi\rangle_{123}=
|\psi^-\rangle_{12}(a|0\rangle-b|1\rangle)_3 \label{123n}
\,.\ea

Clearly, the polarization of the proton target has been teleportated to the outgoing
neutron polarization.
 By a 180$^{\circ}$ rotation around the $y$ direction the state
of the proton beam will be recovered by the outgoing neutron. Note that this experiment
is not complete teleportation because we  discriminate only the  $|\psi^-\rangle $ state\footnote{Contamination
from higher multipole is estimated to be 4\% and it is neglected here.}.

%%%%In Fig. \ref{teleportation}, the layout of an experiment on teleportation of spin states of protons from a
%%%%polarized $^1$H target into the outgoing neutron.
%%%% C is carbon target which operates as an analyzer of the proton polarization using the left-right asymmetry of scattering.
% Proton spin-state is being teleported from the  $^1$H target placed
If the teleportation is taking place in such reaction a strong correlation should be observed
between the polarization of the outgoing neutron and the polarization of the proton beam. Specially, when the polarization of the beam
is in $x$ direction the expected polarization of the outgoing neutron is in $-x$ direction (flipped) however
for  $y$ polarization no flipping should be observed.
Special requirement of the experiment are listed in the table \ref{table1}.
\begin{table}[t]
\caption{\label{table1}Special requirement for the QT experiment with $\vec{d}(\vec{p}$,$^2$He)n  reaction
}
\begin{center}
\begin{tabular}{l|l|l|l|l}\hline\hline
target &beam&$E_d$ &reaction  & detection \\
polarization&polarization&&type&systems
\\\hline
$ P_z\sim 0$ & $a|0\rangle+b|1\rangle$&170 MeV&proton    &$P_{\rm neutron}$   \\
 $ P_{zz}\sim -2$&&&knockout& $E_{\rm neutron}$  \\ \hline
\end{tabular}
\end{center}
\vskip -0.8cm
\end{table}

\section{Conclusion}
 I have shown that the teleprotation of massive Fermion could be implemented using
the available experimental technique. I have suggested   a new physical
effect for the nuclear polarization measurement namely the `quantum-teleportation-effect'.
For the proposed experiment  the expected neutron polarization
 is more or less zero, however taking into account
such effect the neutron polarization  should be enhanced and
 strongly correlated to the beam polarization.
%Generalization to higher dimension and/or mixed
%ensemble is straight forward and will be addressed
Dynamical study that couples the quantum entanglement with the nuclear decoherence
due to the polarization transfer and the induced polarization will be addressed
elsewhere.

%%%%%%%%%%%%%%%%%%%%%%%%%%%%%%%%%%%%%%%%

\end{document}